\documentclass[12pt]{article}
\usepackage{amsmath}
\usepackage{amsfonts}
\usepackage{amssymb}
\usepackage{amsthm}
\usepackage{cite}
\usepackage{hyperref}

\newlength{\dinwidth}
\newlength{\dinmargin}
\newcommand{\kk}{q}
\newcommand{\pp}{\xi}
\newcommand{\gh}{h}
\newcommand{\tih}{\tilde h}
\newcommand{\tig}{\tilde g}
\newcommand{\hf}{\hat f}
\newcommand{\ele}{\mathrm{el}}
\newcommand{\tom}{\wt{\om}}
\newcommand{\nm}{n^-}
\newcommand{\np}{n^+}
\newcommand{\slim}{\mathrm{s}\textrm{-}\lim}
\newcommand{\dGa}{\mathrm{d}\Ga }
\newcommand{\gc}{\la}
\newcommand{\g}{\la }
\newcommand{\ch}{h''}   
\newcommand{\cc}{\mathrm{c} }
\newcommand{\wt}{\widetilde}
\newcommand{\tE}{\tilde{E}}
\newcommand{\trho}{\hat{\rho}}

\newcommand{\ti}{\tilde}

\newcommand{\des}{\mathrm{des}}

\newcommand{\Om}{\Omega}

\newcommand{\Si}{\Sigma}

\newcommand{\tf}{\tilde{f}}
\newcommand{\hh}{h'}   
\newcommand{\ka}{\kappa}

\newcommand{\f}{\mathrm{ph}}
\newcommand{\be}{\beta}
\newcommand{\wh}{\widehat}
\newcommand{\pa}{\partial}

\newcommand{\Ran}{\mathrm{Ran}}
\newcommand{\ov}{\overline}

\newcommand{\mfh}{\mathfrak{h}}

\newcommand{\eps}{\varepsilon}
\newcommand{\de}{\delta}

\newcommand{\De}{\Delta}
\newcommand{\e}{\mathrm{e}}

\newcommand{\pho}{\mathrm{ph}}

\newcommand{\te}{\mathrm}
\newcommand{\nin}{\noindent}
\newcommand{\si}{\sigma}
\newcommand{\ph}{\phantom}
\newcommand{\h}{\fr{1}{2}}
\newcommand{\nat}{\mathbb{N}}
\newcommand{\hil}{\mathcal{H}}
\newcommand{\om}{\omega}
\newcommand{\mfa}{\mathfrak{A}}

\newcommand{\supp}{\mathrm{supp}}
\newcommand{\fr}[2]{\frac{#1}{#2}}
\newcommand{\al}{\alpha}
\newcommand{\real}{\mathbb{R}}
\newcommand{\complex}{\mathbb{C}}
\newcommand{\la}{\lambda}
\newcommand{\non}{\nonumber}
\newcommand{\Ga}{\Gamma}

\def\proof{\noindent{\bf Proof. }}
\def\qed{$\Box$\medskip}

\newtheorem{theoreme}{Theorem } [section]
\newtheorem{proposition}[theoreme]{Proposition}
\newtheorem{lemma}[theoreme]{Lemma}
\newtheorem{definition}[theoreme]{Definition}
\newtheorem{corollary}[theoreme]{Corollary}
\newtheorem{remark}[theoreme]{Remark}
\newtheorem{example}[theoreme]{Example}
\newtheorem{criterion}[theoreme]{Criterion}

\newcommand{\beq}{\begin{equation}}
\newcommand{\eeq}{\end{equation}}
\newcommand{\beqa}{\begin{eqnarray}}
\newcommand{\eeqa}{\end{eqnarray}}
\newcommand{\ben}{\begin{arabicenumerate}}
\newcommand{\een}{\end{arabicenumerate}}
\newcommand{\bex}{\begin{example}}
\newcommand{\eex}{\end{example}}
\newcommand{\ber}{\begin{remark}}
\newcommand{\eer}{\end{remark}}
\newcommand{\bec}{\begin{corollary}}
\newcommand{\eec}{\end{corollary}}
\newcommand{\bep}{\begin{proposition}}
\newcommand{\eep}{\end{proposition}}
\newcommand{\becr}{\begin{criterion}}
\newcommand{\eecr}{\end{criterion}}

\def\bel{\begin{lemma}}
\def\eel{\end{lemma}}
\def\bet{\begin{theoreme}}
\def\eet{\end{theoreme}}
\def\bed{\begin{definition}}
\def\eed{\end{definition}}

\title{Towards a construction of inclusive collision cross-sections in  the massless Nelson model}
\author{
{\bf Wojciech Dybalski\footnote{Supported by the DFG grant SP181/25.}}\\
Zentrum Mathematik, Technische Universit\"at M\"unchen,\\
D-85747 Garching, Germany\\
E-mail: {\tt dybalski@ma.tum.de}}

\begin{document}
\date{ }
\maketitle
\begin{abstract}

The conventional approach to the infrared problem in  perturbative quantum 
electrodynamics relies on the concept of inclusive collision cross-sections. 
A non-perturbative variant of this notion was introduced  in algebraic quantum field theory.
Relying on these insights, we take first steps towards a non-perturbative construction of 
inclusive collision cross-sections in the massless Nelson model. We show that our proposal
is consistent with the standard scattering theory in the absence of the infrared problem
and discuss its status in the infrared-singular case.

\end{abstract}

\section{Introduction}
\setcounter{equation}{0}
The interpretation of physical states of quantum electrodynamics (QED) at asymptotic times is plagued by the 
computational and conceptual difficulties known as the infrared problem \cite{BN37}.   
On the perturbative side a partial solution was given in \cite{JR54, YFS61} with the help of the concept of
inclusive cross-sections. These  cross-sections  incorporate all the outgoing photon configurations whose energy is below the  
sensitivity of the detector. An  attempt to go beyond the inclusive cross-sections was made in \cite{FK70,Ch65, Ki68}, where a concrete expression for scattering states involving  soft-photon clouds (i.e. infinite families of low energy photons) was proposed. 

The two approaches mentioned above  have their counterparts on the non-perturbative side: In the algebraic framework of local relativistic quantum field theory (QFT) a  model independent construction of inclusive cross-sections, proposed in \cite{Bu90, BPS91} and developed further in 
\cite{Jo91,Po04.1,Po04.2,Dy08.3,DT11}, is a promising yet still incomplete programme. In the complementary setting of non-relativistic QED careful choices of soft-photon clouds resulted in a rigorous construction of scattering states for a class of simplified models of QED \cite{Pi03, DG04, Pi05,CFP07,CFP09}, including the Nelson model.

In the present paper we take first steps towards a construction of inclusive cross-sections in the massless Nelson model
along the lines set in algebraic QFT. Let us  therefore recall the  formulation of QED in this latter setting. As usually in relativistic quantum theory, one assumes the  existence of the total energy and momentum operators $(H,P)$, acting on the physical Hilbert space $\hil$, whose joint spectrum is contained in the closed forward lightcone. Moreover, 
 $\hil$ carries the Faraday tensor $F^{\mu\nu}$ and the conserved current $j^{\nu}$, defined as Wightman quantum fields (i.e. certain distributions taking values in unbounded operators on a domain in $\hil$ \cite{SW}),
and subject to the Maxwell equations. 
By smearing the distributions $j^{\nu}$, $F^{\mu\nu}$ with suitable test-functions and taking bounded functions of the resulting operators
one obtains elements of the algebra of observables $\mfa\subset B(\hil)$.

As shown in \cite{Bu86} using the Maxwell equations,  the physical Hilbert space does not contain single-electron states. More precisely, the relativistic mass operator $\sqrt{H^2-P^2}$ has no eigenvectors whose electric charge is different from zero. Hence the electron is an \emph{infraparticle} what precludes the conventional approach to scattering theory based on the LSZ asymptotic fields \cite{Ha58,Ru62,Bu77,Dy05}. An alternative approach, proposed in \cite{Bu90, BPS91}, aims at a direct construction of inclusive cross-sections, 
without recourse to asymptotic fields. 
As a first step one has to identify in the theoretical setting  particle detectors of some finite energy sensitivity $\de$. To this end  one chooses an observable $B_{\de}\in\mfa$ which decreases energy by $\de$ and is almost local in the sense of \cite{AH67}. Next, one introduces the following sequences of observables 
\beq
C_{\de,t}(\chi):= \e^{iHt}\bigg(\int_{\real^3} dx\, \chi(x/t)\, \e^{-iPx}(B^*_{\de}B_{\de})\e^{iPx}\bigg)\e^{-iHt} \label{counters}
\eeq
which have the interpretation, as $t\to\infty$, of particle detectors  sensitive to particles whose velocity is contained in the
support of the function $\chi\in C_0^{\infty}(\real^3)$ and whose energy is larger than $\de$ \cite{AH67}. It was shown in \cite{Bu90} that in any local, relativistic QFT the sequences $\{ C_{\de,t}(\chi)\}_{t\in\real}$ have \emph{weak limit points} $C_{\de}^+$ as 
$t\to\infty$ which are closable operators on a dense domain of states of bounded energy. The structure of these limit points was
investigated in \cite{Po04.1, Po04.2}, conditions for their non-triviality were found in \cite{Dy08.3} and a general scheme
for construction of inclusive collision cross-sections  was proposed in \cite{BPS91}. So far this scheme was  tested only in  asymptotically complete theories of massive    particles, where no infrared problems arise \cite{St89}. More importantly, the  problem of convergence
of the sequences $\{ C_{\de,t}(\chi)\}_{t\in\real}$, which is essential for their  interpretation as particle detectors, remains
open to date in  the presence of infraparticles. This problem is the main obstacle to the understanding of 
infraparticle scattering in the general framework of algebraic QFT. In order to shed some light on this issue, we reformulate
the programme outlined in \cite{BPS91} in the setting of non-relativistic QED, where this and related questions  appear to
be more tractable.  

Let us consider the massless, translationally invariant Nelson model given by the Hamiltonian and momentum operators
\beqa
H&=&\fr{p^2}{2M}+\int_{\real^3}d\kk\, |\kk|a^*_\kk a_\kk+\gc \int_{\real^3}d\kk\fr{\trho(\kk)}{\sqrt{2|\kk|}}(\e^{-i\kk x}a^*_\kk+\e^{i\kk x}a_\kk),\label{Nelson-Hamiltonian}\\
P&=&p+\int_{\real^3} d\kk\, \kk\, a^*_\kk a_\kk, \label{Nelson-momentum}
\eeqa
acting on the Hilbert space $\hil=L^2(\real^3)\otimes \Ga(L^2(\real^3))$. Here the first factor is the space of states of the massive
particle, which we call `electron', whose position and momentum operators are denoted by $(x,p)$. The second factor is the bosonic Fock space   containing states of scalar massless particles, which we call `photons'. The corresponding creation and annihilation operators are denoted by $a^*_\kk,a_\kk$. The form factor $\trho\in C^{\infty}_0(\real^3)$ is the Fourier transform of the charge density $\rho$ of the electron. If the total charge  $(2\pi)^{3/2}\trho(0)$ and the coupling constant $\gc$ 
are different from zero, then the Hilbert space does not contain states describing a single (dressed) massive particle. More precisely, the operator $H-E(P)$, where $\xi\to E(\xi)$ is the lower boundary of the joint spectrum of the family of commuting operators $(H,P)$, has no eigenvectors with eigenvalue zero \cite{HH08,Fr73}. Scattering states in this infraparticle situation were constructed in \cite{Pi03,Pi05} by a careful description of the 
soft-photon clouds accompanying the massive particle. The goal of the  present work  is complementary: We propose a strategy aiming at a direct construction of (inclusive) cross-sections, avoiding a consideration of scattering states. The first step is to identify suitable counterparts of the particle detectors (\ref{counters}) in the present concrete setting. 

To construct a photon detector, we use the following approximants
\beq
C_{g,t}(\chi):=\e^{itH}\dGa(g\chi(y/t)g)\e^{-itH}, \label{photon-velocity}
\eeq   
where $(y,k)$ are the photon position and momentum operators, $\chi, g\in C_0^{\infty}(\real^3)$ are real valued, the support 
of  $g$ is isolated from zero, and we denote $g:=g(k)$. As $t\to\infty$, this observable has an interpretation of a detector
sensitive to photons whose momentum belongs to the support of $g$ and whose direction of motion is restricted by the support 
of $\chi$ (cf. Proposition~\ref{photon-proposition}).  $\{C_{g,t}(\chi)\}_{t\in\real}$ converges strongly to the 
limit $C_g^+(\chi)$ on all vectors from the relevant energy range. This fact, stated as Theorem~\ref{photon-counter} below,  
can be proven using methods developed in \cite{DG99,FGS04}.

As an electron detector we propose the asymptotic velocity of the electron, approximated by
\beq
C_t(f):=\e^{itH}f(x/t)\e^{-itH}, \label{asymptotic-velocity}
\eeq
where $f\in C_0^{\infty}(\real^3)$. At asymptotic times this observable is sensitive to electrons whose velocity belongs to the support of $f$ (cf. Proposition~\ref{electron-proposition}). In the infraparticle situation the convergence of $\{C_t(f)\}_{t\in\real}$ to a particle detector $C^+(f)$ is known only on the infraparticle scattering states constructed in \cite{Pi03,Pi05}. The problem of convergence
of this sequence of operators on all vectors from the relevant energy subspace does not seem to be amenable to existing methods.
This important open problem will not be addressed in the present paper, apart from some remarks in Section~\ref{Outlook}.

Let us now outline the construction of inclusive cross-sections with the help of the particle detectors $C_g^+(\chi)$ and $C^+(f)$, which is the main subject of the present paper. To leave the problem of convergence of $\{C_t(f)\}_{t\in\real_+}$ aside, let us introduce an infrared cut-off i.e. suppose that $\kk\to\trho(\kk)$ vanishes in a neighbourhood of zero. Under this assumption it was shown in \cite{FGS04} that for a sufficiently small coupling constant $\gc$ the model is asymptotically complete up to a certain energy value $\Si>\inf\si(H)$. Restricting attention to the corresponding subspace $\mathbf{1}_{(-\infty,\Si]}(H)\hil$  of the Hilbert space, it is easy to prove the existence of the particle detectors $C_g^+(\chi)$ and $C^+(f)$. 
(See Propositions~\ref{electron-proposition} and \ref{photon-proposition}). Now let $S^+\subset\real^3$ be a compact subset of momenta of the electron. With the help of a sequence of functions $f_n(\nabla E(\,\cdot\,))$, tending pointwise to the characteristic function of $S^+$, we construct $C^+_{S^+}:=\slim_n  C^+(f_n)$. Similarly, for any compact subset $R^+\subset\real^3\backslash \{0\}$ of  photon momenta, the corresponding detector $C^+_{R^+}$ is constructed. Next, for any prescribed configuration of (hard) particles, given by mutually disjoint sets $S^+, R_1^+,\ldots, R_{\np}^+$, we define the set of total energies of this configuration
\beq
\De=\{\,E(\xi)+|\kk_1|+\cdots+|\kk_{\np}|\,|\, \xi\in S^+,\, \kk_1\in R_1^+,\ldots, \kk_{\np}\in R_{\np}^+\, \}. 
\eeq
Then  $|\De|:=\sup \De-\inf \De$ gives the experimental uncertainty of the total energy of the system. The energies of hard
photons should be larger than this uncertainty i.e. $|\De|<\inf\{\, |\kk|\,|\, \kk\in R_j^+, 1\leq j\leq n^+\,\}$ should hold. It is our
main result that under this assumption the operator
\beq
Q^+:=\mathbf{1}_{\De}(H)C^+_{R_1^+}\ldots C^+_{R_{\np}^+}C^+_{S^+}\label{projections}
\eeq   
is an \emph{orthogonal projection} thus it has a clear quantum mechanical interpretation
in terms of particle measurements. Every non-zero vector from the range of this projection describes the prescribed configuration
of hard particles and some unspecified configuration of soft photons whose total energy is less than~$|\De|$. 
(See Theorem~\ref{Projections}).  By reversing the time direction, the above construction can be repeated for 
an incoming configuration of hard particles 
$S^-,R_1^-,\ldots, R_{\nm}^-$ resulting in a projection $Q^-$.  The (inclusive) cross-section for the collision process
$S^-, R_1^-,\ldots, R_{\nm}^-  \to  S^+,R_1^+,\ldots, R_{\np}^+$ is proportional to the transition probability
\beqa
\mathrm{Pr}=\bigg|\bigg(\fr{Q^-\Psi^-}{\|Q^-\Psi^-\|}, \fr{Q^+\Psi^+}{\|Q^+\Psi^+\|}\bigg)\bigg|^2,
\eeqa
where $\Psi^{\pm}\in  \mathbf{1}_{(-\infty,\Si]}(H)\hil$ are chosen so that $Q^{\pm}\Psi^{\pm}\neq 0$. The physical 
situation dictates the choice $\Psi^+=Q^-\Psi^-$. Then, exploiting the fact that $Q^{\pm}$
are orthogonal projections, we obtain
\beqa
\mathrm{Pr}=\bigg(\fr{Q^-\Psi^-}{\|Q^-\Psi^-\|},Q^+ \fr{Q^-\Psi^-}{\|Q^-\Psi^-\|}\bigg). \label{transition-probability}
\eeqa
The remaining dependence of this quantity on the vector $\Psi^-$ reflects the ambiguities inherent in the experimental procedure of the initial state preparation.

While the technical aspects of the above construction still rely on the standard scattering theory, we conjecture that the 
operators $Q^{\pm}$ also exist and are orthogonal projections in the presence of the infrared problem. 
In this general context they should provide a natural language for a description of collision processes, which has a counterpart in the relativistic setting of algebraic QFT  \cite{BPS91}. 

Our paper is organized as follows: In Section~\ref{Preliminaries} we recall some known facts about the Nelson model which are
relevant to the present investigation. In Section~\ref{Particle-counters} the convergence of the particle detector approximants 
(\ref{photon-velocity}) and (\ref{asymptotic-velocity}) is established on the subspace of scattering states. The proofs of these
results, which exploit methods presented in \cite{FGS04}, are given in the Appendices. In Section~\ref{Cross-sections} 
we prove that the operators $Q^+$, given by (\ref{projections}), are orthogonal projections on the subspace of scattering states. In Section~\ref{Outlook} we discuss the feasibility of  our construction in the presence of 
the infrared problem. In particular, we point out that the convergence of photon detectors can be established in this situation.



\section{Preliminaries}\label{Preliminaries}
\setcounter{equation}{0}

In this section, which serves mostly to introduce notation, we provide a survey of some known facts
about the Nelson model, which will be needed in the sequel.
  
Let $\mfh$ be a Hilbert space and let $\Ga(\mfh):=\oplus_{n\geq 0}\otimes_s^n\mfh$ be the symmetric Fock space
over $\mfh$. We denote by $a^*(h)$ and $a(h)$, $h\in\mfh$, the creation and annihilation operators and
set $\phi(h)=a^*(h)+a(h)$. For any self-adjoint operator $b$, acting on  (a dense domain in) $\mfh$, we denote by $\dGa(b)$ a self-adjoint
operator, acting on (a dense domain in)  $\Ga(\mfh)$, defined by 
\beq
\dGa(b)|_{\otimes_s^n\mfh}=\sum_{i=1}^n(\underbrace{ I\otimes\cdots\otimes I}_{i-1}\otimes b \otimes I\otimes\cdots I).
\eeq
For a more detailed exposition of the Fock space combinatorics we refer e.g. to \cite{FGS04}.

Let $\hil_{\ele}=L^2(\real^3)$ be the Hilbert space of the (bare) electron, whose position and momentum operators are denoted by
$(x,p)$. Let $\mfh=L^2(\real^3)$ be the Hilbert space of a single photon, whose  position and momentum operators are denoted by
$(y,k)$. The normalized elements of $\hil_{\ele}$ (resp. $\mfh$) are electron (resp. photon) wave-functions in momentum space. On the full Hilbert space $\hil=\hil_{\ele}\otimes\Ga(\mfh)$ we define the Hamiltonian and momentum operators of
the Nelson model 
\beqa
H&=&\Om(p)+H_{\pho}+\phi(G_x), \label{Nelson-Hamiltonian-one}\\
P&=&p+P_{\pho}. \label{Nelson-Momentum-one}
\eeqa
Here $\Om(p)=\fr{p^2}{2M}\otimes I$, $H_{\pho}=I\otimes \dGa(\om)$, where $\om(q)=|q|$ and $\om:=\om(k)$, $P_{\pho}=I\otimes \dGa(k)$,   
$G_x(\kk)=\ka(\kk)\e^{-i\kk x}$, where  $\ka(\kk)=\g\fr{\trho(\kk)}{\sqrt{2\om(\kk)}}$ and  $\trho\in C_0^{\infty}(\real^3)$ is a positive function.
It is a consequence of the Kato-Rellich theorem that $H$ is self-adjoint on the domain of self-adjointness of $H_0:=\Om(p)+H_{\pho}$ and its spectrum $\si(H)$ is bounded from below.

The operators $(H, P)$, given by (\ref{Nelson-Hamiltonian-one}), (\ref{Nelson-Momentum-one}), form a family of four
commuting self-adjoint operators on a domain in $\hil$. The lower boundary of their joint spectrum  is denoted by $\xi\to E(\xi)$. 
We recall the following facts:
\bel\label{spectral} There exists  $\Si>\inf\,\si(H)$ and  $\gc>0$ s.t. for  $E(\xi)<\Si$ and some $\eps>0$
\begin{enumerate}
\item[(a)] $\xi\to E(\xi)$  is real analytic,
\item[(b)] $\xi\to \nabla E(\xi)$ is invertible,
\item[(c)] $|\nabla E(\xi)|<1$,
\item[(d)] $\||\nabla\Om(p)|\mathbf{1}_{(-\infty,\Si+\eps]}(H)\|<1$.
\end{enumerate}
\eel
Parts (a)-(c) of this lemma follow from \cite{AH10}. Part (d) is a consequence of Lemma~11 of \cite{FGS04}. 
We assume in the sequel that $\Si$ and $\gc$ are chosen as in the above lemma.   

Next, we define the subspace of dressed single-particle states
\beq
\hil_{\des}=\mathbf{1}_{(-\infty, \Si]}(H)\mathbf{1}_{\{0\}}(H-E(P))\hil.
\eeq 
In Sections~\ref{Particle-counters} and \ref{Cross-sections} we assume that $\hil_{\des}\neq \{0\}$.
This holds e.g. in the presence of an infrared cut-off, 
i.e. if $\kk\to \trho(\kk)$ vanishes in some neighbourhood of zero \cite{Fr73, FGS04},
and is expected to hold whenever the massive particle is neutral i.e.
if $\trho(0)=0$. On the other hand, $\hil_{\des}=\{0\}$ if $\g\neq 0$ and  the massive particle 
is charged i.e. $\trho(0)>0$ \cite{Fr73,HH08}. We hope that our construction of inclusive cross-sections  
can be extended to this general situation. We discuss some first steps in Section~\ref{Outlook}.

Proceeding to the construction of asymptotic creation and annihilation operators of massless particles,  
we introduce the spaces
\beqa
L^{2}_{\om}(\real^3)&=&\{\, h\in L^2(\real^3)\,|\, \|h\|_{\om}^2:=\int_{\real^3} d\kk |h(\kk)|^2\big(1+\om(q)^{-1}\big)<\infty\,\},\\
L^{2,\om}(\real^3)&=&\{\, h\in L^2_{\om}(\real^3)\,|\,  \om h\in  L^2_{\om}(\real^3)\, \}.
\eeqa
We recall the following results:
\bel\label{power-bounds} (a) For every  $n\in\nat$ the operator $H_\f^n(H+i)^{-n}$ is bounded.\\ 
(b) For every $n\in\nat$ there is a constant $C_n$, s.t. for all $h_1,\ldots, h_n\in L_{\om}^2(\real^3)$ 
\beq
\|a^{\#}(h_1)\ldots a^{\#}(h_n)(H_\f+1)^{-n/2}\|\leq C_n\prod_{j=1}^N \|h_j\|_{\om}, 
\eeq
where $a^{\#}(\,\cdot\,)$ stands for a creation or annihilation operator.
\eel 
Part (a) of this lemma is an adaptation of an argument from \cite{FGS01} to the case of the Nelson model, 
part (b) is standard. In view of the above lemma, we can define the asymptotic creation operator
\beq
a^*_{+}(h)\Psi:=\slim_{t\to\infty}\e^{itH}a^*(h_t)\e^{-itH}\Psi,
\eeq
where $h_t=\e^{-i\om t}h$ and $\Psi$ is any vector from the domain of $(c+H)^{\h}$ for which the limit exists. 
(Here $c\geq 0$ is s.t. $\si(H)+c\subset (0,\infty)$). The following fact, which is a result from \cite{GZ09} adapted to the case of the Nelson model, concerns the properties of these operators:
\bel\label{existence-of-ac} Choose  $\Si>\inf\,\si(H)$ and  $\gc>0$ s.t. $\||\nabla\Om(p)|\mathbf{1}_{(-\infty,\Si+\eps]}(H)\|<1$ for some 
$\eps>0$. (Cf. Lemma~\ref{spectral}).
Then, for any  $h_1,\ldots, h_n\in L^{2,\om}(\real^3)$ 
and  $\Psi\in\mathbf{1}_{(-\infty,\Si]}(H)\hil$ the following  limit 
\beq
\slim_{t\to\infty}\e^{itH}a^{\#}(h_{1,t})\ldots a^{\#}(h_{n,t})\e^{-itH}\Psi \label{asymptotic-creation-one}
\eeq
exists and  equals 
\beq
a_+^{\#}(h_{1})\ldots a_+^{\#}(h_{n})\Psi.\label{asymptotic-creation-two}
\eeq
\eel
Let us now prove the following lemma, which is a sharpened variant of  Theorem~4~(iv) of \cite{FGS01}:
\bel\label{harmonic-analysis} Let $\Psi\in \Ran\mathbf{1}_{[E_1,E_2]}(H)$, where $E_1\leq E_2\leq \Si$. 
Let  $h_1,\ldots, h_n\in  L^{2,\om}(\real^3)$ be functions
of compact support (in the sense of distributions), $m_j:=\inf\{\,\om(\kk)\,|\,\kk\in\supp\,h_j\,\}$ and 
$M_j:=\sup\{\,\om(\kk)\,|\,\kk\in\supp\,h_j\,\}$. Then the vector
\beq
\Psi_n:=a_+^{*}(h_{n})\ldots a_+^{*}(h_{1})\Psi
\eeq
 belongs to $\Ran\,\mathbf{1}_{[E_1+m_1+\cdots+m_n,E_2+M_1+\cdots+M_n]}(H)$.
\eel
\proof We proceed by induction. For $n=0$ the statement is trivially true. Suppose it holds for
$n-1$ creation operators. We choose $\hf\in C_0^{\infty}(\real)$ s.t. $\hf(\om(\,\cdot\,))=1$ on $\supp\,h_n$ and 
vanishes on a slightly larger set. We write 
\beqa
& &a_+^{*}(h_{n})\Psi_{n-1}=a_+^{*}(\hf(\om)h_{n})\Psi_{n-1}\non\\
& &=\slim_{t\to\infty}\fr{1}{\sqrt{2\pi}}\int ds\, f(s) \e^{iHt}a^*(\e^{-i\om t}\e^{i\om s}h_n)\e^{-iHt}\Psi_{n-1}\non\\
& &=\slim_{t\to\infty}\fr{1}{\sqrt{2\pi}}\int ds\, f(s) \e^{iHs}\big(\e^{iHt}a^*(\e^{-i\om t}h_n)\e^{-iHt} \big)\e^{-iHs}\Psi_{n-1},
\eeqa
where in the last step we made use of the dominated convergence theorem for Bochner integrable functions to enter with the limit under the integral sign and make a
change of variables. (Norm continuity of the integrand follows from Lemma~\ref{power-bounds}). Next, we recall that for any $A\in B(\hil)$ and compact set $\De\subset \real$ 
the operator $A(f):=\int ds\, f(s)\e^{iHs}A\e^{-iHs}$ satisfies \cite{Ar82}
\beq
A(f)\mathbf{1}_{\De}(H)\hil\subset \mathbf{1}_{\{\ov{\De+\supp\,\tf}\}}(H)\hil.\label{energy-transfer}
\eeq 
Each operator $A:=\e^{iHt}a^*(\e^{-i\om t}h_n)\e^{-iHt}\mathbf{1}_{\De_{n-1}}(H)$, where $\De_{n-1}=[E_1+m_1+\cdots+m_{n-1},E_2+M_1+\cdots+M_{n-1}]$, belongs to $B(\hil)$ by Lemma~\ref{power-bounds}. Making use of the induction hypothesis and of formula~(\ref{energy-transfer}), we complete the proof. \qed\\
Let $L^{2,\om}_{\cc}(\real^3)$ be the subspace of  functions from $L^{2,\om}(\real^3)$ which are compactly 
supported (in the sense of distributions). We  define the following space 
\beqa
\hil_0^+=\te{Span}\{\, a_+^*(h_1)\ldots a_+^*(h_n)\Psi\,|\, \Psi\in\hil_{\des}, h_j\in  L^{2,\om}_{\cc}(\real^3), 1\leq j\leq n, n\in\nat \,\}\quad\,\,\, \label{hilplus}  
\eeqa
as well as its norm closure $\hil^+=(\hil_0^+)^{\mathrm{cl}}$. 


\section{Particle detectors}\label{Particle-counters}
\setcounter{equation}{0}
In this section we collect the relevant properties of the electron detectors $C^+(f)$ and the photon detectors $C_{g}^+(\chi)$, 
defined as limits of the respective sequences
\beqa
C_t(f)&:=&\e^{itH}f(x/t)\e^{-itH}, \label{asymptotic-velocity-1}\\
C_{g,t}(\chi)&:=&\e^{itH}\dGa(g\chi(y/t)g)\e^{-itH}, \label{photon-velocity-1}
\eeqa
as $t\to\infty$. Here $f\in C_0^{\infty}(\real^3)$, $\chi, g\in C_0^{\infty}(\real^3)_{\real}$, the support of $g$ does not
contain zero and we wrote $g:=g(k)$ in (\ref{photon-velocity-1}) above. In particular, we show that these limits exist on vectors from $\hil_0^+$.

The following proposition describes the action of the electron detector on scattering states. 
Although this result may be known (cf. Remark 2i in \cite{Pi05}), we have not found a proof in the literature.
Thus we include the straightforward argument in Appendix~\ref{Electron-counter}. 
\bep\label{electron-proposition}  Let $h_1,\ldots, h_n\in  L^{2,\om}_{\cc}(\real^3)$, $f\in C_0^{\infty}(\real^3)$ and 
$\Psi\in\hil_{\des}$. Then 
\beq
\slim_{t\to\infty}\e^{iHt}f(x/t)\e^{-iHt}a_+^*(h_1)\ldots a_+^*(h_n)\Psi=a_+^*(h_1)\ldots a_+^*(h_n)
f(\nabla E(P))\Psi. \label{AH-electron}
\eeq
\eep 

The action of the photon detector on scattering states is described in the following proposition, proven in Appendix~\ref{Photon-counter}.
\bep\label{photon-proposition} Let $\eps>0$ be s.t. $|\nabla E(\xi)|<1-\eps$ for $E(\xi)\leq \Si$. (Cf.~Lemma~\ref{spectral}). 
Choose $\chi\in C_0^\infty(\real^3)_{\real}$,
s.t. $\supp\,\chi \subset\{\, v\in\real^3\,|\, ||v|-1|\leq \eps/4\,\}$ and $\chi(v)=1$ if $|v|=1$. 
Let $h_1,\ldots, h_n\in  L^{2,\om}_{\cc}(\real^3)$  
and $\Psi\in\hil_{\des}$. Then 
\beqa
& &\slim_{t\to\infty}\e^{iHt}\dGa\big(g\chi(y/t)g\big)\e^{-iHt}a_+^*(h_1)\ldots a_+^*(h_n)\Psi\non\\
& &\ph{44444444}=\sum_{j=1}^n a_+^*(h_1)\ldots a_+^*\big( g^2h_j\big)\ldots a_+^*(h_n)\Psi \label{AH-photon}
\eeqa  
for $g\in C_0^{\infty}(\real^3)_{\real}$ vanishing in a neighbourhood of zero. 
\eep

The above results justify the physical interpretation of  $C^+(f)$, $C_{g}^+(\chi)$, as particle detectors
sensitive to particles whose momenta are specified by the supports of the functions $f$ and $g$.
For the construction of collision cross-sections  it is essential that these operators
are invariant under time translations. While it is evident on $\hil_0^+$ by the above two propositions, it
can easily be shown in more generality.
\bep\label{translation-invariance} Let $\Si'>0$ be arbitrary and  let $\Psi_1,\Psi_2\in \mathbf{1}_{(-\infty,\Si']}(H)\hil$ be s.t.
the limits
\beqa
\e^{isH}C^+(f)\e^{-isH}\Psi_1&:=&\slim_{t\to\infty} \e^{i(t+s)H}f(x/t)\e^{-i(t+s)H}\Psi_1,\\
\e^{isH}C^+_g(\chi)\e^{-isH}\Psi_2&:=& \slim_{t\to\infty} \e^{i(t+s)H}\dGa(g\chi(y/t)g)\e^{-i(t+s)H}\Psi_2
\eeqa
exist for  $s=0$. Then the limits exist for all $s\in\real$ and $\e^{isH}C^+(f)\e^{-isH}\Psi_1=C^+(f)\Psi_1$, $\e^{isH}C^+_g(\chi)\e^{-isH}\Psi_2=C^+_g(\chi)\Psi_2$.
\eep
\nin The proof of this proposition, which partly comes from \cite{FGS04}, is given in Appendix~\ref{invariance-of-counters}.



\section{Inclusive collision cross-sections} \label{Cross-sections}

\setcounter{equation}{0}
In this section we give a detailed construction of the operators $Q^{+}$, defined precisely in (\ref{Q+def}) below, 
which we used to define transition probabilities of inclusive collision processes in (\ref{transition-probability}). In particular,
we show that these operators are orthogonal projections on $\hil^+$, what validates
their interpretation as coincidence arrangements of particle detectors.

Let $S^+\subset\real^3$ be a compact subset of electron momenta s.t. $E(\xi)\leq \Si$ for $\xi\in S^+$. 
Making use of Lemma~\ref{spectral} (b), we choose a family of  functions
$f_{\eps}\in C_0^{\infty}(\real^3)$ s.t. $\xi\to f_{\eps}(\nabla E(\xi))$ approximates pointwise the 
characteristic function $\mathbf{1}_{S^+}(\,\cdot\,)$ of the set $S^+$ as $\eps\to 0$. Exploiting Proposition~\ref{electron-proposition} and Lemma~\ref{power-bounds},
we can meaningfully define for any $h_1,\ldots,h_n\in L^{2,\om}_{\cc}(\real^3)$ and $\Psi\in\hil_{\des}$   
\beqa
C^+_{S^+} a_+^*(h_1)\ldots a_+^*(h_n)\Psi
 &:=&\slim_{\eps\to 0}\slim_{t\to\infty}\e^{iHt}f_{\eps}(x/t)\e^{-iHt}a_+^*(h_1)\ldots a_+^*(h_n)\Psi\non\\
&=&a_+^*(h_1)\ldots a_+^*(h_n)\mathbf{1}_{S^+}(P)\Psi.
\eeqa
Clearly, $C^+_{S^+}$ extends to an orthonormal projection on $\hil^+$.  

Now let $R^+\subset \real^3$ be a compact subset of photon momenta s.t. $R^+\cap \{0\}=\emptyset$. 
Let $\kk\to g_{\eps}(\kk)$ be a family of positive functions from $C_0^{\infty}(\real^3)$, vanishing in some neighbourhood of zero, which tends pointwise to the characteristic function $\mathbf{1}_{R^+}(\,\cdot\,)$ of $R^+$ and let $\chi$ be chosen as in 
Proposition~\ref{photon-proposition}. Making use of  Proposition~\ref{photon-proposition} and Lemma~\ref{power-bounds}, we define   
\beqa
& &C^+_{R^+} a_+^*(h_1)\ldots a_+^*(h_n)\Psi\!\!\!\non\\
& &\ph{44444}:=\slim_{\eps\to 0}\slim_{t\to\infty}\e^{iHt}\dGa(g_{\eps}\chi(y/t)g_{\eps})\e^{-iHt}a_+^*(h_1)\ldots a_+^*(h_n)\Psi\non\\
& &\ph{444444444444444444444}=\sum_{j=1}^n a_+^*(h_1)\ldots a_+^*(\mathbf{1}_{R^+}h_j)\ldots a_+^*(h_n)\Psi.
\eeqa
$C^+_{R^+}$ is an unbounded operator. It is defined on $\hil^+_0\subset \hil^+$ and leaves this subspace invariant.

Next, given a collection of sets $S^+,R^+_1,\ldots,R^+_n$, we define the set of total energies of the corresponding particle configuration
\beq
\De=\{\, E(\pp)+\om(\kk_1)+\cdots+\om(\kk_n)\,|\, \pp\in S^+, \kk_j\in R^+_j, 1\leq j\leq n \,\}. \label{Del}
\eeq
Then $|\De|:=\sup \De-\inf \De$ is the accuracy with which the total energy of this particle configuration is  known. 
The sets $R^+_1,\ldots,R^+_n$ should describe `hard photons', whose energies are larger than $|\De|$. 
If this condition is met, then the corresponding operator $Q^{+}$, given by (\ref{Q+def}) below, is an orthogonal projection, 
what validates its interpretation as a quantum mechanical measurement. This is the content of the following theorem, which
is our main result.  
\bet\label{Projections} Let $S^+,R^+_1,\ldots,R^+_n$ be compact subsets of $\real^3$  s.t.   $R^+_i\cap R^+_j=\emptyset$ for $i\neq j$ 
and $E(\xi)\leq \Si$ for $\xi\in S^+$. Let $\De$ be given by (\ref{Del}).
If $|\De|<\inf\{\, \om(\kk)\,|\, \kk\in R^+_j, 1\leq j\leq n\, \}$, then the operator
\beq
Q^+:=\mathbf{1}_{\De}(H)C^+_{R^+_1}\ldots C^+_{R^+_n} C^+_{S^+} \label{Q+def}
\eeq
is an orthogonal projection on $\hil^+$. Moreover, $\Ran\, Q^+$ is spanned by vectors of the form
\beqa
a_+^*(h_1)\ldots a_+^*(h_n)a_+^*(\hh_{n+1})\ldots a_+^*(\hh_m)\Psi,
\eeqa
where $m\geq n$, $\Psi\in\mathbf{1}_{S^+}(P)\hil_{\des}$, $h_i,\hh_j\in L^{2,\om}_{\cc}(\real^3)$, $\supp\, h_i\subset R^+_i$, 
$\supp\,\hh_j\subset\{\,\kk\in\real^3 \,|\, |\kk|\leq \de_j  \,\}$, $\de_{n+1}+\cdots+\de_m\leq |\De|$. 
\eet
\proof The subspace $\hil^+_0$ is spanned by vectors of the form 
\beq
\Psi^+=a_+^*(h_1)\ldots a_+^*(h_m)\Psi,
\eeq
where $h_i\in L^{2,\om}_{\cc}(\real^3)$, $1\leq i\leq m$ and $\Psi\in\hil_{\des}$. 
It is clear from Proposition~\ref{photon-proposition} that $Q^+\Psi=0$ for $m<n$. 
For $m\geq n$ the expression $Q^+\Psi^+$ is a sum of terms of the form
\beqa
\Psi^+_1:=\mathbf{1}_{\De}(H)a_+^*(\mathbf{1}_{R^+_{i_1}}h_1)\ldots a_+^*(\mathbf{1}_{R^+_{i_n}} h_n)
 a_+^*(h_{n+1})\ldots a_+^*(h_m)\mathbf{1}_{S^+}(P)\Psi.\label{many-massless-photons}
\eeqa 
We decompose $h_{n+i}=\hh_{n+i}+\ch_{n+i}$, where $\supp\,\ch_{n+i}\subset \{\, k\in\real^3 \,|\,|k|\geq |\De|\,\}$, 
$\supp\,\hh_{n+i}\subset \{\,k\in\real^3 \,|\,  |k|<|\De|\,\}$ and $1\leq i\leq (m-n)$. Then, $Q^+\Psi^+$ is a sum of terms of the form
\beqa
\Psi^+_2:=\mathbf{1}_{\De}(H)a_+^*(\mathbf{1}_{R^+_{i_1}}h_1)\ldots a_+^*(\mathbf{1}_{R^+_{i_n}} h_n)a_+^*(\ch_{n+1})\ldots a_+(\ch_{n+j})\non\\
\ph{4444444444444444444}\cdot a_+^*(\hh_{n+j+1} )\ldots a_+^*(\hh_m) \mathbf{1}_{S^+}(P)\Psi. \label{range-of-the-projection}
\eeqa
We will show that terms for which $j\geq 1$ are zero. First, we note that
\beq
E_{\te{max}}:=\sup \De=\sup \,\{ E(\xi)\,|\,\xi\in S^+\}+\sum_{j=1}^n \sup\{\,\om(\kk)\,|\, \kk\in R^+_j\,\}.
\eeq
Next, we obtain from Lemma~\ref{harmonic-analysis} that  $\Psi^+_2\in \Ran\,\mathbf{1}_{[E_{\te{min}},\infty)}(H)$, where 
$E_{\min}$ satisfies 
\beq 
E_{\te{min}}\geq  \inf\,\{ E(\xi)\,|\,\xi\in S^+\}+\sum_{j=1}^n \inf\{\,\om(\kk)\,|\, \kk\in R^+_j\,\}+|\De|.   
\eeq
Hence $E_{\min}-E_{\te{max}}\geq 0$. Since $\Psi^+_2$ is not an eigenvector of $H$,  $\Psi^+_2=0$ for $j\geq 1$.
We conclude that $Q^+\Psi^+$ is a sum of terms of the form
\beq
\Psi^+_2:=\mathbf{1}_{\De}(H)a_+^*(\mathbf{1}_{R^+_{i_1}}h_1)\ldots a_+^*(\mathbf{1}_{R^+_{i_n}} h_n)a_+^*(\hh_{n+1})\ldots a_+^*(\hh_{m}) \mathbf{1}_{S^+}(P)\Psi. \label{sample-vector}
\eeq
Since $C^+_{R^+_i}$, $C^+_{S^+}$ commute with $H$ by Proposition~\ref{translation-invariance},  $\mathbf{1}_{R^+_{i}}\hh_j=0$ by assumption and $\mathbf{1}_{R^+_{i'}}\mathbf{1}_{R^+_{j'}}=0$ for $i'\neq j'$, it is evident that $Q^+Q^+\Psi^+=Q^+\Psi^+$.  

To prove the second part of the theorem, it suffices to justify the support property of the functions $h_j'$. It follows from the above discussion that vectors of the
form (\ref{sample-vector}) span $\Ran\, Q^+$. Let us fix $\eps>0$ and let $l=(m-n)$ be the number
of soft photons in  $\Psi^+_2$. It suffices to consider the case $l>0$. We define $\chi_j$ to be the characteristic function of the set 
$\{\, \kk\in\real^3 \,|\,  (j-1)\fr{\eps}{l}\leq |\kk|\leq j  \fr{\eps}{l}  \}$, where $j\in\{1,\ldots, [ l\fr{|\De |}{\eps} ]+1\}$. 
Then $\hh_{n+i}(\kk)=\sum_{j_{n+i}} \chi_{j_{n+i}}(\kk)\hh_{n+i}(\kk)$. Substituting this decomposition to (\ref{sample-vector}), we
obtain that $\Psi^+_2$ is a sum of terms of the form
\beqa
\Psi^+_3&:=&\mathbf{1}_{\De}(H)a_+^*(\mathbf{1}_{R^+_{i_1}}h_1)\ldots a_+^*(\mathbf{1}_{R^+_{i_n}} h_n)\non\\
& &\ph{4444444444444}\cdot a_+^*(\chi_{j_{n+1}}\hh_{n+1} )\ldots a_+^*(\chi_{j_{m}}\hh_{m}) \mathbf{1}_{S^+}(P)\Psi. 
\eeqa
If the above vector is non-zero, then Lemma~\ref{harmonic-analysis} gives
\beq
(j_{n+1}-1)\fr{\eps}{l}+\cdots+(j_{m}-1)\fr{\eps}{l}\leq |\De|.
\eeq
Now let $\de_{n+i}:=\sup\{\, |\kk| \,|\, \kk\in\supp\, \chi_{j_{n+i}}\hh_{n+i}\,\}$. Since $\de_{n+i}\leq j_{n+i}  \fr{\eps}{l}$,
it follows from the above relation that $\de_{n+1}+\cdots+\de_m\leq |\De|+\eps$. By choosing $\hat f_n\in S(\real)$ s.t.
$\slim_{n\to\infty} \hat f_n(H)=\mathbf{1}_{\De}(H)$, writing $\hat f_n(H)=\fr{1}{\sqrt{2\pi}}\int dt\, \e^{iHt} f_n(t)$ and
making use of the fact that $\e^{iHt}a_+^*(h) \e^{-iHt}=a_+^*(\e^{i\om t} h)$, we obtain the following:
For any $\eps>0$, $Q^+\Psi^+$ belongs to the subspace $\hil_{\eps}$ spanned by vectors of the form
\beqa
a_+^*(h_1)\ldots a_+^*(h_n)a_+^*(\hh_{n+1})\ldots a_+^*(\hh_m)\Psi,
\eeqa
where $m\geq n$, $\Psi\in\mathbf{1}_{S^+}(P)\hil_{\des}$, $h_i,\hh_j\in L^{2,\om}_{\cc}(\real^3)$, $\supp\, h_i\subset R^+_i$, $\supp\,\hh_j\subset\{\,k\in\real^3 \,|\, |k|\leq \de_j  \,\}$, $\de_{n+1}+\cdots+\de_m\leq |\De|+\eps$. 
By identifying $\hil_{\eps}$ with a subspace of
$\Ga(L^2(\{\, k\in \real^3\,|\, |k|\geq |\De|\,\}))\otimes \Ga(L^2(\{\, k\in \real^3\,|\, |k|\leq |\De|\,\}))\otimes\hil_{\des}$ and noting
that 
\beqa
\slim_{\eps\to 0} I\otimes  \mathbf{1}_{[0,|\De|+\eps]} (H_{\pho})\otimes I
=I\otimes  \mathbf{1}_{[0,|\De|]} (H_{\pho})\otimes I,
\eeqa
we obtain that $\bigcap_{\eps>0}\hil_{\eps}$ coincides with the subspace described in the statement of the theorem. \qed\\
The construction of the subspace of incoming scattering states $\hil^-$ and of the corresponding projections $Q^-$ 
proceeds analogously as above, by taking the limit $t\to-\infty$.  Assuming that $\mathbf{1}_{(-\infty, E]}(H)\hil^+=\mathbf{1}_{(-\infty, E]}(H)\hil^-$ for some $E\leq \Si$, (which holds under conditions specified in \cite{FGS04} as a consequence of asymptotic completeness) transition probabilities for inclusive collision processes below this energy can be defined using formula~(\ref{transition-probability}).

\section{Conclusion and outlook}\label{Outlook}

\setcounter{equation}{0}
In this work we made first steps towards a construction of inclusive collision
cross-sections in the massless Nelson model, following  ideas developed in 
algebraic quantum field theory \cite{BPS91}. We identified suitable asymptotic
observables which play the role of particle detectors. Their coincidence arrangements 
can be used for preparation of incoming and outgoing configurations of hard particles
accompanied by some unspecified configurations of soft photons.

We tested the proposed construction in the absence of the infrared problem, to show that 
it is consistent with predictions of standard scattering theory. 
As for the infrared-singular case, the first question is the existence of the particle detectors $C_g^+(\chi)$
and $C^+(f)$. In the case of the photon detector, a positive answer can
be inferred from the existing literature: Using methods developed in \cite{DG99}, the convergence 
of $t\to\e^{iHt}\dGa(\chi(y/t))\e^{-iHt}$
is established in \cite{FGS04} for the Nelson model with an infrared cut-off. The convergence of 
$t\to\e^{iHt}\dGa(g\chi(y/t)g)\e^{-iHt}$, where $g$ vanishes in a neighbourhood of
zero, can be established analogously in the absence of the infrared cut-off. 
Thus, by a minor modification of the proof of Theorem~26 of \cite{FGS04},
one obtains:
\bet\label{photon-counter} Let $\g, \Sigma$ be s.t. $\|\,|\nabla\Om(p)| \mathbf{1}_{(-\infty,\Si]}(H)\|\leq\be$
for some $\be<1$. Let $\chi\in C_0^{\infty}(\real^3)_{\real}$ be s.t. 
$\supp\,\chi\subset\{\, v\in\real^3\,|\, |v|>\be\,\}$ and suppose that $g\in C_0^\infty(\real)_{\real}$ vanishes in a 
neighbourhood of zero. Then  the limit 
\beq
C_g^+(\chi)\Psi:=\slim_{t\to\infty}\e^{iHt}\dGa(g\chi(y/t)g)\e^{-iHt}\Psi
\eeq
exists for any $\Psi\in\Ran\mathbf{1}_{(-\infty,\Si)}(H)$.  
\eet
The problem of existence of $C^+(f)=\slim_{t\to\infty}\e^{iHt}f(x/t)\e^{-iHt}$
in the presence of the infrared problem appears to be  more difficult. To our knowledge, it is only
resolved on the subspace of infraparticle scattering states, constructed in \cite{Pi03,Pi05}. We
recall that an analogous asymptotic observable (asymptotic velocity) plays a central role in quantum mechanical scattering theory,
where it exists under very general conditions \cite{DG}. There the main ingredient of the 
proof of convergence is the equality of the average velocity $x/t$ and the instantaneous velocity $p/M$
of the particle at asymptotic times (Graf's propagation estimate \cite{Gr90}). This route seems
difficult  in the setting of non-relativistic QED due to the phenomenon of the electron
mass renormalization. We hope, however, that  recently developed powerful time dependent \cite{DK10} and spectral \cite{CFFS09} methods  
will shed some light on the question of existence of the  asymptotic velocity 
of the electron in models of non-relativistic QED. 
In view of the framework proposed in the present paper,  an answer to this question will 
allow for a meaningful definition of inclusive collision cross-sections in these models.
It may also clarify the problem of convergence of  the  particle detector approximants~(\ref{counters})
and thus contribute to the understanding of the infrared problem  in relativistic (algebraic) quantum field theory.

\bigskip
\bigskip
\bigskip
\bigskip
\bigskip
\bigskip

\noindent{\bf Acknowledgements:}
I would like to thank  D.~Buchholz, J. Derezi\'nski,  A.~Pizzo,  W.~De Roeck and H.~Spohn for interesting
discussions on scattering theory. The hospitality extended to me at final stages of this work by the 
University of Heidelberg and the University of California, Davis is gratefully acknowledged.

\appendix


\section{Proof of Proposition~\ref{electron-proposition} }\label{Electron-counter}

\setcounter{equation}{0}
In this Appendix we give a proof of Proposition~\ref{electron-proposition}. 
More technical part of this discussion is postponed to subsequent lemmas.\\
\bf Proof of Proposition~\ref{electron-proposition}: \rm  Making use of Lemma~\ref{harmonic-analysis}, 
we obtain
\beqa
& &\e^{iHt}f(x/t)\e^{-iHt}a_+^*(h_1)\ldots a_+^*(h_n)\Psi\non\\
& &\ph{4444444444444444444}=\e^{iHt}f(x/t)\e^{-iHt}g'(H)a_+^*(h_1)\ldots a_+^*(h_n)\Psi
\eeqa
for some $g'\in C^{\infty}_0(\real)$. There holds
\beqa
& &\e^{iHt}f(x/t)g(H)a^*(h_{1,t})\ldots a^*(h_{n,t})\Psi_{t}=\e^{iHt}[f(x/t),g'(H)]a^*(h_{1,t})\ldots a^*(h_{n,t})\Psi_{t}\non\\
& &\ph{44444444444444444444444444}+\e^{iHt}g'(H) a^*(h_{1,t})\ldots a^*(h_{n,t})f(x/t)\Psi_{t},\label{AH-main-equality}
\eeqa
where $\Psi_{t}=\e^{-iHt}\Psi$.
The first term on the r.h.s. tends to zero as $t\to\infty$ by Lemmas~\ref{commutator-f-g} and \ref{power-bounds}.
To treat the second term, we choose a real-valued function  $\tE\in C^{\infty}_0(\real^3)$ which coincides with
$\xi\to E(\xi)$ for $E(\xi)\leq\Si$.  
We write
\beqa
& &\e^{iHt}g'(H) a^*(h_{1,t})\ldots a^*(h_{n,t})f(x/t)\Psi_{t}\non\\
&=&\e^{iHt}g'(H) a^*(h_{1,t})\ldots a^*(h_{n,t})\e^{-i\tE(P)t} \big(\e^{i\tE(P)t} f(x/t)\e^{-i\tE(P)t}\Psi-f(\nabla \tE(P))\Psi\big)\non\\
&+&\e^{iHt}g'(H) a^*(h_{1,t})\ldots a^*(h_{n,t})\e^{-iHt}f(\nabla \tE(P))\Psi.
\eeqa
Since $\|g'(H) a^*(h_{1,t})\ldots a^*(h_{n,t})\|$ is bounded uniformly in time, (see Lemma~\ref{power-bounds}),
the first term on the r.h.s. above converges to zero, by Lemma~\ref{resolvent-convergence} with the above choice of the function $\tE$. The second term converges 
to the r.h.s. of (\ref{AH-electron}) by Lemma~\ref{harmonic-analysis}. This concludes the proof. \qed\\
In the proof of the auxiliary Lemma~\ref{commutator-f-g}, stated below, there enter two ingredients. The first is a simple fact from the pseudodifferential
calculus, whose proof can be found in \cite{FGS01}. 
\bel\label{Pseudodifferential}  Let $f\in S(\real^3)$, $\gh\in C^n(\real^3)$ and $\sup_{|\al|=2}\|\pa^\al \gh\|_{\infty}<\infty$,
where $\pa^\al=\pa_1^{\al_1}\pa_2^{\al_2}\pa_3^{\al_3}$ and $|\al|=\al_1+\al_2+\al_3$. Then
\beqa
i[\gh(k),f(\eps x)]&=&i\eps\nabla \gh(k)\cdot \nabla f(\eps x)+R_{1,t}(\eps)\non\\
&=& i\eps \nabla f(\eps x)\cdot \nabla \gh(k)+R_{2,t}(\eps),
\eeqa
where
\beq
\|R_{j,t}(\eps)\|\leq C\eps^2 \sup_{|\al|=2}\|\pa^{\al}\gh\|_{\infty}\int_{\real^3} du\, |u|^2 |\hf(u)|
\eeq
for some constant $C$ independent of $\eps$.
\eel
The second ingredient is the Helffer-Sj\"ostrand functional calculus, which we summarize following \cite{FGS01}:
Let $f\in C_0^\infty(\real,\complex)$ and $A$ be a self-adjoint operator. Then $f(A)$ can be represented as follows
\beq
f(A)=-\fr{1}{\pi}\int du dv\,  \pa_{\bar z}\tilde f(z)(z-A)^{-1},\quad z=u+iv.
\eeq
This holds for any function $\tf\in C_0^\infty(\real^2,\complex)$ s.t. $|\pa_{\bar z}\tf(z)|\leq C|v|$, $\tf(z)=f(z)$ and 
$\pa_{\bar z}\tilde f(z)=\fr{1}{2}\big(\pa_u f+i\pa_w f\big)(z)=0$   
for all  $z\in\real$. Such $\tf$ is called an almost-analytic extension of $f$. For any $n\in \nat$ there exist extensions for 
which $|\pa_{\bar z}\tf(z)|\leq C|v|^n$. 
\bel\label{commutator-f-g} For any $f\in C_0^{\infty}(\real^3)$ and $\gh\in C_0^{\infty}(\real)$ there holds
\beq
 \|[\gh(H),f(x/t)]\|\leq C/t
\eeq 
for some constant $C\geq 0$ independent of $t$. 
\eel
\proof We choose an almost-analytic extension of $\gh$ s.t $|\pa_{\bar z}\ti \gh(u,v)|\leq C|v|^2$. Then, for $z=u+iv$
\beqa
& &[\gh(H),f(x/t)]\non\\
& &=-\fr{1}{\pi}\int dudv\,\pa_{\bar z}\ti \gh(u,v)[(z-H)^{-1}, f(x/t)]\non\\
& &=-\fr{1}{\pi}\int dudv\,\pa_{\bar z}\ti \gh(u,v)(z-H)^{-1}[\Om(p),f(x/t)](z-H)^{-1}.
\eeqa
Now by Lemma~\ref{Pseudodifferential},  $[\Om(p),f(x/t)]=\fr{1}{t}(\nabla f)(x/t)\nabla\Om(p)+R_2(t)$, where $\|R_2(t)\|\leq c'/t^2$.
We note the following estimate
\beq
\|\pa_i\Om(p)(z-H)^{-1}\|\leq \|\pa_i\Om(p)(c+H)^{-\h}\|\,\|(c+H)^{\h}(z-H)^{-1}\|, \label{two-factors}
\eeq
where $c\geq 0$ is chosen s.t. $\si(H)+c\subset (0,\infty)$.
The first factor on the r.h.s. of (\ref{two-factors}) above can be estimated as follows:
\beqa
\|\pa_i\Om(p)(c+H)^{-\h}\|^2&=&\|(c+H)^{-\h}|\pa_i\Om(p)|^2(c+H)^{-\h}\|\non\\
&\leq& (2/M)\|(c+H)^{-\h}(\Om(p))(c+H)^{-\h}\|\non\\
&\leq&(2/M)\|(H+c_1)(c+H)^{-1}\|,
\eeqa
where the bound $\Om(p)\leq H+c_1$ for some constant $c_1$ follows e.g. from Lemma~8 of \cite{FGS04}. 
The second factor on the r.h.s. of (\ref{two-factors}) gives 
\beqa
\|(c+H)^{\h}(z-H)^{-1}\|^2&=&\|(\bar z+H)^{-1}(c+H)(z+H)^{-1}\|\non\\
&=&\sup_{w\geq 0}\fr{w}{(u+w-c)^2+v^2} \non\\
&=&\sup_{w\geq R}\fr{w}{(u+w-c)^2+v^2} \non\\
&+&\sup_{0\leq w\leq R}\fr{w}{(u+w-c)^2+v^2}, \label{u-v-decomposition}
\eeqa
where $R\geq \sup\{\,|u|\,|\, \pa_{\bar z}\ti g(u,\,\cdot\,)\equiv 0  \}+c+1$. The second term on the r.h.s. 
of (\ref{u-v-decomposition}) can be estimated by $R/v^2$ and the first one by $\sup_{w\geq R} w/(w-R+1)^2\leq C$. Hence,
\beq
\|\pa_i\Om(p)(z-H)^{-1}\|\leq C/|v|.
\eeq
Making use of the  above facts and of the obvious inequality $\|(z-H)^{-1}\|\leq C/|v|$,  
we complete the proof. \qed\\
Let us now proceed to the second auxiliary lemma which we used in the proof of Proposition~\ref{electron-proposition}.
\bel\label{resolvent-convergence} Let $f\in C_0^{\infty}(\real^3)$ and $\tE\in C_0^{\infty}(\real^3)$ be a real-valued function. Then   
\beq
\slim_{t\to\infty}\e^{i\tE(P)t}f(x/t)\e^{-i\tE(P)t}=f(\nabla \tE(P)). \label{AH-states}
\eeq  
\eel
\proof Let $\Psi\in\hil$.  By the functional calculus for a family of commuting self-adjoint operators, we obtain 
\beq
\e^{i\tE(P)t}f(x/t)\e^{-i\tE(P)t}\Psi=f(x/t+\nabla\tE(P))\Psi.
\eeq
We note  that, by Theorem~VIII.25 of \cite{RS1}, $x^i/t+\pa_i \tE(P)$ converges to $\pa_i \tE(P)$ in the strong
resolvent sense. In fact, the domain of $x^i$, denoted $D(x^i)$, is a common core for all the approximants and the limit,
and for any $\Phi\in D(x^i)$ 
\beq
\slim_{t\to\infty}(x^i/t+\pa_i \tE(P))\Phi=\pa_i\tE(P)\Phi.
\eeq
Thus we obtain (\ref{AH-states}) e.g. by approximating $f$ in the supremum norm by functions of the form 
$f_{\eps}(x)=\sum_{j} f_{\eps,j,1}(x^1)f_{\eps,j,2}(x^2)f_{\eps,j,3}(x^3)$, where $f_{\eps,j,k}\in S(\real)$ and the sum is finite. \qed


\section{Proof of Proposition~\ref{photon-proposition} }\label{Photon-counter}
\setcounter{equation}{0}
In this Appendix we prove Proposition~\ref{photon-proposition} which establishes
the existence of the photon detectors on scattering states of bounded energy. As in Appendix~\ref{Electron-counter}, the more technical
part of this discussion is given in subsequent lemmas. \\
\bf Proof of Proposition~\ref{photon-proposition}: \rm
By Lemma~\ref{harmonic-analysis}, we obtain
\beqa
& &\e^{iHt}\dGa(g\chi(y/t)g)\e^{-iHt}a_+^*(h_1)\ldots a_+^*(h_n)\Psi\non\\
& &\ph{44444}=\e^{iHt}\dGa(g\chi(y/t)g)\e^{-iHt}g'(H)a_+^*(h_1)\ldots a_+^*(h_n)\Psi\non\\
& &\ph{44444}=\e^{iHt}[\dGa(g\chi(y/t)g),g'(H)]\e^{-iHt}a_+^*(h_1)\ldots a_+^*(h_n)\Psi\non\\
& &\ph{44444}+\e^{iHt} g'(H) \dGa(g\chi(y/t)g) \e^{-iHt}a_+^*(h_1)\ldots a_+^*(h_n)\Psi, \label{main-photon-AH}
\eeqa 
for some $g'\in C_0^{\infty}(\real)$. It follows from Lemma~\ref{photon-lemma} and Proposition~\ref{electron-proposition} that the term with 
the commutator tends to zero. As for the last term on the r.h.s. of (\ref{main-photon-AH}), 
we note that 
\beq
\sup_{t}\|g'(H)\dGa(g\chi(y/t)g)\|<\infty. \label{uniformity}
\eeq
This is a consequence of Lemmas~\ref{dGa-bound} and \ref{power-bounds}~(a).
Thus we can write
\beqa
& &g'(H)\e^{iHt} \dGa(g\chi(y/t)g) \e^{-iHt}a_+^*(h_1)\ldots a_+^*(h_n)\Psi\non\\
& &\ph{44444}=g'(H)\e^{iHt} \dGa(g\chi(y/t)g)a^*(h_{1,t})\ldots a^*(h_{n,t})\Psi_{t}+o(1)\non\\
& &\ph{44444}=g'(H)\e^{iHt}\sum_{j=1}^n a^*(h_{1,t})\ldots a^*(g\chi(y/t)gh_{j,t})\ldots a^*(h_{n,t})\Psi_{t}\non\\
& &\ph{44444}+g'(H)\e^{iHt}a^*(h_{1,t})\ldots a^*(h_{n,t}) \dGa(g\chi(y/t)g) \Psi_{t}+o(1),\label{many-terms-AH}
\eeqa
where $o(1)$ denotes a rest term which tends to zero in norm as $t\to\infty$ and we used that 
$[\dGa(g\chi(y/t)g), a^*(h_{j,t})]=a^*(g\chi(y/t)gh_{j,t})$.
Let us first study one of the terms in the sum above. We will show that $a^*(g\chi(y/t)gh_{j,t})$ can be replaced with
$a^*(\{g^2\chi(\nabla\om)h_j\}_t)$ at a cost of an error term of order $o(1)$. We set 
\beq
\tih_{j,t}:=\e^{-i\om t}g\big(\e^{i\tom t}\chi(y/t)\e^{-i\tom t}-\chi(\nabla\tom)\big)gh_j,
\eeq
where $\tom$ is a smooth, compactly supported function which coincides with $\om$ on the support of $g$.
We obtain
\beqa
& &\|g'(H)\e^{iHt}a^*(h_{1,t})\ldots a^*(\tih_j)\ldots a^*(h_{n,t})\Psi_{t}\|\non\\
&=&\|g'(H)\e^{iHt}a^*(h_{1,t})\ldots\check j \ldots a^*(h_{n,t})a^*(\tih_{j,t})\Psi_{t}\|\non\\
&\leq& \|g'(H)\e^{iHt}a^*(h_{1,t})\ldots\check j \ldots a^*(h_{n,t})\|\, \|a^*(\tih_{j,t})g'(H)\|\,\|\Psi\|,
\eeqa
where $\check j$ denotes the omission of the $j$-th creation operator.
By Lemma~\ref{power-bounds} the first factor on the r.h.s. above is bounded uniformly in time, whereas  
the second factor satisfies
\beq
\|a^*(\tih_{j,t})g'(H)\|
\leq C\|(1+\om^{-1})^\h\tih_{j,t}\|,
\eeq
for some constant $C\geq 0$. Denoting $\tig(\kk)=(1+\om(\kk)^{-1})^\h g(\kk)$, we obtain
\beq
\|(1+\om^{-1})^\h\tih_{j,t}\|\leq \|\tig\|_{\infty}\|\big(\e^{i\tom t}\chi(y/t)\e^{-i\tom t}-\chi(\nabla\tom)\big)gh_j\|.
\eeq
Proceeding as in the proof of Lemma~\ref{resolvent-convergence}, we obtain
\beq
\slim_{t\to\infty}\e^{i\tom t}\chi(y/t)\e^{-i\tom t}=\chi(\nabla\tom).\label{strong-convergence-all}
\eeq
Hence, we have shown that
\beqa
& &\sum_{j=1}^n g'(H)\e^{iHt}a^*(h_{1,t})\ldots a^*(g\chi(y/t)gh_{j,t})\ldots a^*(h_{n,t})\Psi_{t}\non\\
&=&\sum_{j=1}^n g'(H)\e^{iHt}a^*(h_{1,t})\ldots a^*(\{g^2\chi(\nabla\tom)h_j\}_t)\ldots a^*(h_{n,t})\Psi_{t}+o(1).\quad\,\,\,\,
\eeqa
Making use of Lemma~\ref{harmonic-analysis} and exploiting the fact that $|\nabla\tom(q)|=1$ for $q\in\supp\,g$, we obtain in the limit 
$t\to\infty$ the r.h.s. of~(\ref{AH-photon}). 

It still has to be shown that the remaining terms on the r.h.s. of (\ref{many-terms-AH}) tend to zero.  
To this end, we write 
\beqa
& &\|g'(H)\e^{iHt}a^*(h_{1,t})\ldots a^*(h_{n,t}) \dGa(g\chi(y/t)g) \Psi_{t}\|\non\\
& &\ph{4444444}\leq\|g'(H)\e^{iHt}a^*(h_{1,t})\ldots a^*(h_{n,t})\|\,\|\dGa(g\chi(y/t)g) \Psi_{t}\|.
\eeqa
Here the first factor on the r.h.s. is uniformly bounded by Lemma~\ref{power-bounds}. The second factor vanishes for $t\to\infty$
by Lemma~\ref{detector-vanishing}.\qed\\   
In the above proof we used the following two lemmas:
\bel\label{photon-lemma} Let $f\in C_0^{\infty}(\real^3)$  be s.t. $\supp\, f\subset \{\,u\in\real^3\,|\, |u|\leq\be\,\}$. 
Choose $\chi\in C_0^\infty(\real^3)$ s.t. $\supp\,\chi \subset\{\, v\in\real^3\,|\, |v|\geq \be+\eps\,\}$
for some $\be,\eps>0$. Then 
\beqa
& &\|[\dGa(g\chi(y/t)g),g'(H)]\|\leq C,\label{uniform-bound}\\
& &\|[\dGa(g\chi(y/t)g),g'(H)]f(x/t)\|\leq C/t,\label{tends-to-zero}
\eeqa  
for $c$ independent of $t$, $g'\in C_0^{\infty}(\real)$ and $g\in C_0^{\infty}(\real^3)_{\real}$ s.t. 
$g$ vanishes  in a neighbourhood of zero. 
\eel
\proof We apply the Helffer-Sj\"ostrand functional calculus (see Appendix~\ref{Electron-counter}). Choosing an almost-analytic 
extension $\ti g'$ of $g'$ s.t. $|\pa_{\bar z}\ti g'(u,v)|\leq C|v|^3$, we obtain
\beqa 
& &i[\dGa(g\chi(y/t)g),g'(H)]\non\\
& &=-\fr{1}{\pi}\int du dv\, \pa_{\bar z}\ti g'(u,v) (z-H)^{-1}\dGa(gi[\tom,\chi(y/t)]g)(z-H)^{-1}\non\\
& &\ph{4}+\fr{1}{\pi}\int du dv\, \pa_{\bar z}\ti g'(u,v) (z-H)^{-1}\phi(ig\chi(y/t)gG_x)(z-H)^{-1}, \label{dGa-inequality}
\eeqa
where $z=u+iv$, $\tom$ is a smooth, compactly supported function which coincides with $\om$ on the support of $g$.
To obtain (\ref{dGa-inequality}), we used
\beqa
i[\dGa(g\chi(y/t)g),\phi(G_x)]&=&\phi(ig\chi(y/t)g G_x),\\ 
i[\dGa(g\chi(y/t)g),\dGa(\om)]&=&\dGa(i[g\chi(y/t)g,\om]).
\eeqa
Next, we denote by $C_{1,t}$, $C_{2,t}$ the first and the second term on the r.h.s. of (\ref{dGa-inequality}), respectively. 
First, we show that
\beq
\|C_{1,t}\|\leq C/t. \label{C_one} 
\eeq
To this end, we note that for $c\geq 0$ s.t.  $\si(H)+c\subset (0,\infty)$ 
\beqa
& &(c+H)^{-\h}\dGa(gi[\tom,\chi(y/t)]g)(c+H)^{-\h}\non\\
& &\ph{444444444444} \leq\fr{C}{t}(c+H)^{-\h}\dGa(g^2)(c+H)^{-\h}+O(t^{-2})\non\\
& &\ph{444444444444} \leq\fr{C'}{t}(c_1H+c_2)(c+H)^{-1}+O(t^{-2}),
\eeqa
where we made use of the pseudodifferential calculus (cf. Lemma~\ref{Pseudodifferential}) and of the fact that 
$\dGa(b_1)\leq \dGa(b_2)$ for any self-adjoint operators $b_1$, $b_2$ s.t. $b_1\leq b_2$. The term  $O(t^{-2})$ denotes
a family of bounded operators s.t. $\|O(t^{-2})\|\leq Ct^{-2}$ for some constant $C$.
Noting, as in the proof of  Lemma~\ref{commutator-f-g}, that $\|(c+H)^\h(z-H)^{-1}\|\leq C/|v|$, we obtain (\ref{C_one}).

As for $C_{2,t}$, we first show that it is bounded uniformly in $t$. To this end, we note that, by Lemma~\ref{power-bounds}~(b),
\beqa
& &\|(1+H_\f)^{-\h}\phi(ig\chi(y/t)gG_x)(1+H_\f)^{-\h}\|\non\\
& &\ph{4444444444444}\leq C\sup_{x}\|(1+\om^{-1})^{\h}g\chi(y/t)gG_x\|\non\\
& &\ph{4444444444444}\leq C\|(1+\om^{-1})^{\h}  g\|\|\chi\|_{\infty}\|g\|_{\infty}\|G\|. \label{phi-estimate}
\eeqa
Exploiting the fact that $\|(1+H_\f)^\h(z-H)^{-1}\|\leq c/|v|$, we obtain that $\|C_{2,t}\|<\infty$ uniformly in $t$. 
This concludes the proof of (\ref{uniform-bound}). To verify (\ref{tends-to-zero}), we still have to 
check that $\|C_{2,t}f(x/t)\|\leq C/t$ as $t\to\infty$. Setting $h_x:=ig\chi(y/t)gG_x$, we obtain
\beqa
& &(z-H)^{-1}\phi(h_x)(z-H)^{-1}f(x/t)\non\\
& &\ph{44444444444444}=(z-H)^{-1}\phi(h_x)(z-H)^{-1}[\Om(p),f(x/t)](z-H)^{-1}\non\\
& &\ph{44444444444444}+(z-H)^{-1}\phi(h_x)f(x/t) (z-H)^{-1}. \label{field-estimate}
\eeqa
By Lemma~\ref{Pseudodifferential}, $i[\Om(p),f(x/t)]=\fr{1}{t}\nabla f(x/t)\nabla\Om(p)+R_t$,
where $\|R_t\|\leq C/t^2$. We note that $\||\nabla\Om(p)|(c+H)^{-\h}\|<\infty$ and $\|(c+H)^\h(z-H)^{-1}\|\leq C/|v|$. 
From the latter inequality and from~(\ref{phi-estimate}) we obtain $\sup_{x}\|(z-H)^{-1}\phi(h_x)(z-H)^{-1}\|\leq C/|v|^2$. Hence the contribution to $C_{2,t}$,
corresponding to the first term on the r.h.s. of (\ref{field-estimate}), is bounded by $C/t$. To estimate the second term on the r.h.s. of
(\ref{field-estimate}), we note that
\beqa
& &\|(1+H_\f)^{-\h}\phi(h_x)f(x/t)(1+H_\f)^{-\h}\|\non\\
& &\ph{444444444444444}\leq C\|(1+\om^{-1})^{\h}g\|_{\infty}\sup_{|x|/t\leq \be}\|\chi(y/t)gG_x\|.
\eeqa
To show that this expression tends to zero, we recall that $G_x(\kk)=\ka(\kk)\e^{-i\kk x}$, where $\ka(\kk)=\g\fr{\trho(\kk)}{\sqrt{2\om(\kk)}}$, 
$\trho\in C^{\infty}_0(\real^3)$ and obtain, similarly  as in Lemma~9 of \cite{FGS04}, 
\beqa
\sup_{|x|/t\leq \be }\|\chi(y/t)gG_x\|^2&\leq& \sup_{|x|/t\leq \be} \int_{|y|/t\geq \be+\eps} dy\, 
|\wh{g\ka}(y-x)|^2\non\\
&\leq& \bigg(\fr{1}{\eps t}\bigg)^{n}\int dy\,| \wh{g\ka}(y)|^2|y|^{n}.
\eeqa 
The integral on the r.h.s. is finite for any $n\geq 0$, since $g\ka\in C_0^{\infty}(\real^3)$. 
Using the estimate $\|(1+H_\f)^\h(z-H)^{-1}\|^2\leq c/|v|^2$, we conclude that  $\|C_{2,t}f(x/t)\|\leq C/t$ as $t\to\infty$. \qed
\bel\label{detector-vanishing} Let $\eps>0$ be s.t. $|\nabla E(\xi)|<1-\eps$ for $E(\xi)\leq \Si$. (Cf. Lemma~\ref{spectral}). 
Then, for any  $\chi\in C_0^\infty(\real^3)$ 
s.t. $\supp\,\chi \subset\{\, v\in\real^3\,|\, ||v|-1|\leq \eps/4\,\}$ 
and $\Psi\in\hil_{\des}$, 
\beq
\lim_{t\to\infty}\|\dGa(g\chi(y/t)g) \Psi_{t}\|=0,
\eeq  
where $\Psi_{t}=\e^{-iHt}\Psi$ and
 $g\in C_0^{\infty}(\real^3)_{\real}$ vanishes in a neighbourhood of zero. 
\eel
\proof We choose a real-valued function $\tE\in C^{\infty}_0(\real^3)$ which coincides with 
$\xi\to E(\xi)$ for $E(\xi)\leq\Si$ and satisfies   $|\nabla \tE(\xi)|<1-\eps$ for all $\xi\in\real^3$.
As $\Psi$ has bounded energy, we can choose $g'\in C_0^{\infty}(\real)$  s.t. $\Psi=g'(H)\Psi$. Since,
 by Lemma~\ref{photon-lemma} and Proposition~\ref{electron-proposition}, $[\dGa(g\chi(y/t)g),g'(H)]\Psi_{t}$ tends 
strongly to zero as $t\to\infty$, we obtain 
\beqa
\|\dGa(g\chi(y/t)g) \Psi_{t}\|=\|g'(H)\e^{i\tE(P)t}\dGa(g\chi(y/t)g)\e^{-i\tE(P)t}\Psi\|+o(1)& &\non\\
\leq c\|\e^{i\tE(P)t}(1+H_\f)^{-\h}\dGa(g\chi(y/t)g)(1+H_\f)^{-\h}\e^{-i\tE(P)t}\Psi_{\f}\|+o(1),& &
\eeqa
where we set $\Psi_{\f}=(1+H_\f)^{\h}\Psi$ and the constant $c$ is independent of $t$. 
We decompose $\Psi_{\f}=\sum_m \Psi_{\f,m}$ into components with fixed photon
number and write
\beqa
\|\e^{i\tE(P)t}(1+H_\f)^{-\h}\dGa(g\chi(y/t)g)(1+H_\f)^{-\h}\e^{-i\tE(P)t}\Psi_{\f}\|^2\ph{444444444}& &\non\\
=\sum_m\|(1+H_\f)^{-\h}\sum_{j=1}^m g(k_j)\chi(y_j/t+\nabla \tE(P))g(k_j)
(1+H_\f)^{-\h}\Psi_{\f,m}\|^2,& & \label{last-step-AH}
\eeqa 
where we identified $\Psi_{\f,m}$ with (classes of) square-integrable functions on $\real\times \real^{3m}$ and made use of the fact that 
$\e^{i\tE(P)t}y_j^i \e^{-i\tE(P)t}=y_j^i+\pa_i\tE(P)t$ on a suitable dense domain in $L^2(\real^3)\otimes L^2(\real^3)^{\otimes m}$. 
In order to enter with the limit $t\to\infty$ under the sum, we  show that the above sequence can be estimated by
a uniformly convergent sequence independent of $t$ and make use of the dominated convergence theorem. To this end, 
we estimate in the $m$-particle subspace
\beqa
& &(1+H_\f)^{-\h}\sum_{j=1}^m g(k_j)\chi(y_j/t+\nabla \tE(P))g(k_j)(1+H_\f)^{-\h}\non\\
& &\leq \|\chi(y/t+\nabla \tE(P))\|\, \|g(k)^2|k|^{-1}\| \sum_{j=1}^m \om(k_j) (1+H_\f)^{-1}\non\\
& &\leq C (1+H_\f)^{-1}H_{\f}\leq C'.
\eeqa
Clearly, $C'$ can be chosen uniformly in $t$ and $m$. Thus we can enter with the limit under the sum in (\ref{last-step-AH}). The expression tends to zero because $y_j/t+\pa_i\tE(P)$ tends to $\pa_i E(P)$ in the strong resolvent sense and $\nabla \tE(P)$  has its spectrum outside of the support of $\chi$. (Cf. the proof of Lemma~\ref{resolvent-convergence}). \qed

\section{ Proof of Proposition~\ref{translation-invariance} }\label{invariance-of-counters}

\setcounter{equation}{0}
\bf Proof of Proposition~\ref{translation-invariance}: \rm  The invariance of $C^+(f)$ follows from the the estimate
\beqa
\|f(x/(t-s))-f(x/t)\|&=&\|\int_0^s\fr{d}{ds'}f(x/(t-s'))ds'\|\non\\
&\leq& \fr{s}{(t-s)}\sup_{v\in\real^3}|\nabla f(v)\cdot v|, 
\eeqa
where the r.h.s. tends to zero as $t\to\infty$. 

The invariance of $C^+_g(\chi)$ can be proven  by an analogous argument (cf. the proof of Theorem 26 of \cite{FGS04}):
We note that
\beqa
\|\dGa\big(g(\chi(y/(t-s))-\chi(y/t))g\big)(c+H)^{-1}\|&\leq& c'\|\int_0^s\,ds'\,\fr{d}{ds'}\chi(y/(t-s'))\|\non\\
&\leq& c'\fr{s}{(t-s)}\sup_{u\in\real^3} |\nabla\chi(u)\cdot u|,  
\eeqa
where $c$ is sufficiently large and in the first step we made use of Lemma~\ref{power-bounds}~(a) and  
Lemma~\ref{dGa-bound} stated below. \qed\\  
The above proof relies on the following lemma:
\bel\label{dGa-bound} Let $\chi$ be a bounded operator on the single-photon space. Then
\beq
\|\dGa(g\chi g)(1+H_{\f})^{-1}\|\leq c\|\chi\|,
\eeq 
where $c$ is independent of $\chi$ and $g\in C_0^\infty(\real)_{\real}$ vanishes in a neighbourhood of zero.
\eel
\proof Without loss of generality we can assume that $\chi$ is self-adjoint. We  note the following
\beqa
\dGa(g\chi g)(1+H_{\f})^{-1}&=&[\dGa(g\chi g),(1+H_{\f})^{-1/2}](1+H_{\f})^{-1/2}\non\\
&+&(1+H_{\f})^{-1/2}\dGa(g\chi g)(1+H_{\f})^{-1/2}.
\eeqa
The second term on the r.h.s. above satisfies
\beq
(1+H_{\f})^{-1/2}\dGa(g\chi g)(1+H_{\f})^{-1/2}\leq  \|\chi\|\, \|\om(k)^{-1} g^2(k)\| H_{\f}(1+H_{\f})^{-1}.
\eeq
As for the first term, we recall that
\beqa
(1+H_{\f})^{-1/2}=\fr{1}{\pi}\int_0^{\infty}d\la\, \la^{-\h}\fr{1}{\la+1+H_{\f}}. 
\eeqa
Thus we obtain
\beqa
& &i[\dGa(g\chi g),(1+H_{\f})^{-1/2}]\non\\
&=&\fr{1}{\pi}\int_0^{\infty}d\la\, \la^{-\h}
(\la+1+H_{\f})^{-1}\dGa(g i[\chi,\tom]g)(\la+1+H_{\f} )^{-1}\non\\
&\leq& \|[\chi,\tom]\|\,\|\om(k)^{-1}g^2(k)\| \fr{1}{\pi}\int_0^{\infty}d\la\, \la^{-\h}(\la+1+H_{\f})^{-1}H_{\f}
(\la+1+H_{\f} )^{-1}\non\\
&\leq& \|[\chi,\tom]\|\,\|\om(k)^{-1}g^2(k)\| \fr{1}{\pi}\int_0^{\infty}d\la\, \la^{-\h}(\la+1)^{-1}, \label{last-equation}
\eeqa
where $\tom$ is a smooth, compactly supported function which coincides with $\om$ on the support of $g$. 
Since the integral on the r.h.s. of (\ref{last-equation}) is convergent, the proof of the lemma is complete. \qed

\end{document}